# Mid-Infrared Modulation of Quantum Emitters in Hexagonal Boron Nitride

Karin Yamamura[1,2], Xinyang Yu[1,2], Chaohao Chen[1,2], Mehran Kianinia[1,2], Christophe Galland[3], Igor Aharonovich[1,2]*

[1] School of Mathematical and Physical Sciences, University of Technology Sydney, Ultimo, New South Wales 2007, Australia
[2] ARC Centre of Excellence for Transformative Meta-Optical Systems, University of Technology Sydney, Ultimo, New South Wales 2007, Australia
[3] Institute of Physics and Center for Quantum Science and Engineering, Ecole Polytechnique Fédérale de Lausanne (EPFL), 1015 Lausanne, Switzerland

* To whom correspondence should be addressed: I.A. Igor.Aharonovich@uts.edu.au

**Abstract**
*Single photon emitters (SPEs) are promising building blocks for practical devices in quantum technologies. Traditionally, these systems are excited using off-resonant visible light through their phonon transitions, yet this process remains poorly understood. Here, we explore the interaction of mid-infrared (MIR) excitation on the properties of SPEs in hexagonal boron nitride. Notably, we present a reversible, non-destructive method to enhance emission from blue SPEs using MIR co-excitation. By resonantly driving defect-localized in-plane infrared-active optical phonon modes near 7.3 um, the MIR field modulates carrier dynamics through a phonon-assisted recombination. This unique feature, not observed previously for defects in solids, is a promising reservoir in a growing toolkit to modulate quantum emitters at room temperature for their use in practical quantum technologies.*

Luminescent defects in wide-bandgap materials that act as single photon emitters (SPEs) are an invaluable resource for applications in quantum technologies[1-3]. A particular feature of most point defects in solids is a narrowband zero phonon line (ZPL) that corresponds to the *n=0* transition from its electronic excited to the ground state, along with a broader vibronic phonon side band (PSB)[4-6]. The coupling of electronic transition to optical phonon branches, facilitates the non-resonant excitation of the SPEs even at room temperature. However, it also introduces non-coherent relaxation pathways (into the PSB), reducing the fraction of emission into the ZPL[7, 8]. In parallel, coupling to acoustic photons leads to dephasing and spectral diffusion that result in a decreased coherence of the ZPL[9, 10]. Therefore, controlling or modulating the local vibrational environment may offer an avenue for enhancing the brightness of the studied point defects while maintaining their intrinsic spectral properties.

In particular, mid-infrared (MIR) irradiation at the phonon resonance can be an efficient tool to modulate the local phonon occupancy. It has recently been utilised to achieve bright upconversion luminescence from nanoparticles and utilised in surface-enhanced Raman scattering and charge stabilisation of quantum dot emission[11-14]. Although the fundamental mechanisms are not universal and are predominantly dependent on the material, these developments motivate the exploration of MIR co-excitation in the context of SPEs in solids.

In this work, we demonstrate a reversible interaction between a MIR laser and a solid state SPE. By resonantly driving lattice vibrations coupled to the SPE, the MIR field modifies carrier population dynamics through a phonon-assisted recombination, resulting in an increased emission intensity.

As our test platform we employ SPEs in hexagonal boron nitride (hBN)[15, 16], and in particular the recently discovered B centre[17-20]. The B centres can be engineered on demand using electron beams[18, 19] and has a consistent zero-phonon line (ZPL) at 436 nm. This is an important advantage to study interaction with a MIR irradiation as similar emitters can be reliably reproduced. In addition, the B centres are coherent[21], and have already been utilised for the generation of indistinguishable photons[22] as well as exploring cavity - coupled systems[23].

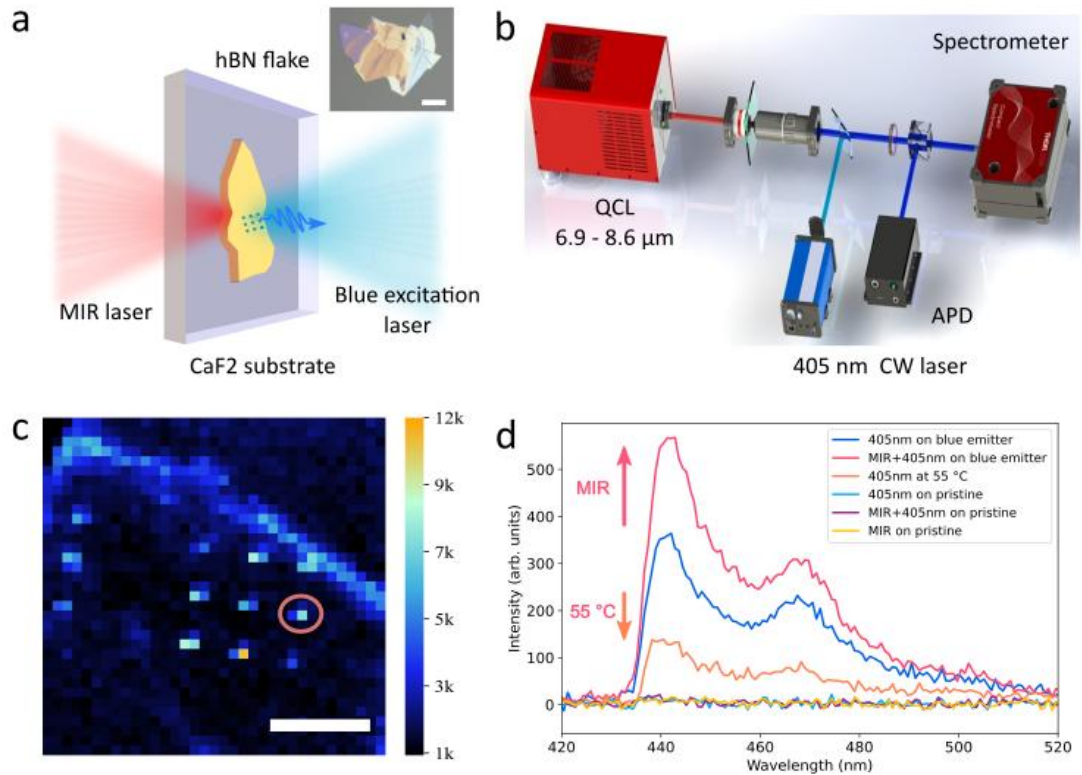

*Figure 1. Experimental configuration and mid-IR co-excitation of a blue single-photon emitter in hBN. (a) Schematic of a B centre in hBN excited by a 405 nm laser, with MIR irradiation applied from the opposite side. The inset shows an optical image of the hBN flake (scale bar: 100 μm). (b) Schematic of the experimental setup for blue emitter characterisation under 405 nm continuous-wave (CW) excitation and mid-IR co-excitation. A quantum cascade laser (QCL, 6.9–8.6 μm) is aligned counter-propagating with respect to the 405 nm excitation path. Emission from the hBN flake on a CaF$_2$ substrate is split by a 50:50 beam splitter fibre and directed to a spectrometer and an avalanche photodiode (APD). (c) Confocal PL map of an electron-beam-irradiated array in an hBN flake. The orange circle marks the emitter measured in (d) (scale bar: 10 μm). (d) PL spectra of the blue emitter ensembles under 405 nm CW excitation alone at room temperature and at 55 °C, and under simultaneous mid-IR co-excitation at 7.3 μm. No PL signal is observed from pristine hBN regions under the same excitation conditions. The acquisition time is 1s.*

Figure 1(a) illustrates the setup configuration of the B-centre in hBN under co-excitation with a 405 nm primary excitation laser and a MIR co-irradiation laser. The inset displays an optical image of an exfoliated hBN flake with a thickness of 50-150 nm. Figure 1(b) shows the experimental setup used for MIR co-excitation of B-centres in hBN. The hBN flake is placed onto a CaF$_2$ substrate and is mounted along the main excitation and collection optical paths. The emitted signal from the B centre was collected through two evenly split detection paths using a 50:50 beam splitter, and directed either to the spectrometer or to an avalanche photo diode. The MIR radiation was generated through a quantum cascade laser (QCL) with operational wavelength from 6.9 μm to 8.6 μm and focused through the back of the sample using a reflective objective (see Methods section for further details).

Figure 1(c) presents a PL confocal map acquired using only the blue excitation laser (405 nm), revealing arrays of the B centres in hBN activated by electron-beam irradiation. The emission from a particular emitter highlighted by the orange circle is shown in Fig. 1(d) as a blue curve. In addition, Fig. 1(d) demonstrates the effect of MIR co-excitation on an individual B centre. MIR co-excitation is resonant with a phonon mode of the emitter ($\hbar\omega_{MIR} \approx \hbar\omega_{phonon}$), where $\omega_{MIR}$ is the MIR laser frequency and $\omega_{phonon}$ is the optical phonon mode ($E_{1u}$) frequency of bulk hBN at *7.3μm (~1370cm$^{-1}$)*[24, 25]. A pronounced enhancement in PL intensity is observed under MIR co-excitation at milliwatt-level power (red curve). The MIR co-excitation increases the occupation probability of the emissive state by reshaping carrier relaxation dynamics via electron-phonon coupling. This process promotes the depopulation of trapped carriers and returns them to either the ground or excited state, thus enhancing radiative emission without introducing measurable changes in the emission spectrum (as will be discussed further below). In contrast, no detectable emission is observed from either the emitter or the surrounding hBN flake under MIR excitation alone (yellow and purple curves). Similarly, no PL signal is detected when identical excitation conditions are applied to pristine regions of the hBN flake (light blue curve).

To exclude the possibility that the observed PL enhancement under MIR co-excitation could originate from local heating, we performed temperature-dependent measurements up to 55 °C using a home-built heating stage. The sample temperature was monitored with a thermocouple attached to the edge of the heating platform. The PL spectra acquired at 55 °C (orange curve) exhibit an overall decrease in emission intensity compared to that at room temperature (blue curve). To further analyse the temperature-induced spectral changes, the PL spectra were fitted using Gaussian functions to extract the emission characteristics of the ZPL at 436 nm and three phonon sidebands located at 443 nm, 462 nm, and 480 nm, labelled PSB1, PSB2, and PSB3, respectively. The assignment of these spectral features follows the identification established in previous studies based on spectral fitting of blue emitters in hBN[26]. Table SI1 and SI2 in the supplementary information list the centre wavelengths, FWHM, and maximum intensities of the ZPL and three PSBs extracted from the fitted data. The fitting reveals a slight red-shift and linewidth broadening of the ZPL and the PSB1 at elevated temperature. It is important to note that there's no such change in the spectra under the MIR co-excitation (see Table SI3 and SI4).

Such temperature-induced PL quenching and spectral broadening are well-known signatures of enhanced electron–phonon interactions due to lattice expansion in hBN-based emitters[27, 28]. These thermal effects and emission quenching are not observed in our experiments under MIR co-excitation. This strongly indicates that the MIR-induced PL enhancement cannot be attributed to heating and instead arises from a non-thermal, wavelength-selective excitation mechanism.

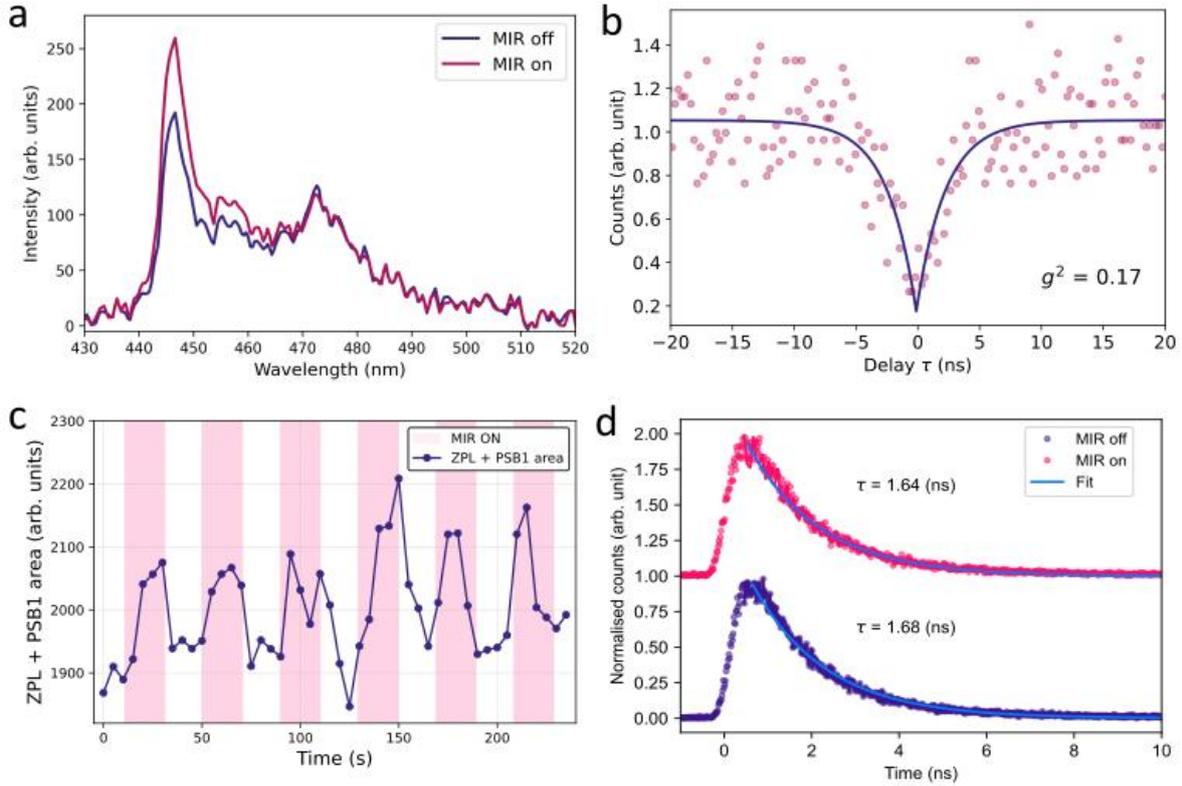

*Figure 2. MIR modulation of the B centre and blue emitter ensembles. (a) PL spectra measured from a B centre with (red curve) and without (blue curve) MIR radiation at 7.3 μm, or ~1370 cm$^{-1}$. (b) Second-order autocorrelation function, $g^{(2)}(\tau)$, recorded from the same emitter. (c) Time trace of the area under the main PL peak from a blue emitter ensemble showing reversible enhanced emission during the MIR on state (shaded in pink). (d) Lifetime comparison with and without the MIR field measured from the same emitter in (c). The plots were offset for clarity.*

To confirm that the MIR-induced PL enhancement acts on single quantum emitters, we repeat the MIR excitation on individual SPEs. Figure 2(a) shows the excitation of a single B centre with (purple) and without (blue) MIR co-irradiation. In both cases, the same excitation wavelength of 405 nm and the same laser power were used. Figure 2(b) shows a clear single-photon characteristic, with a second-order autocorrelation value of $g^{(2)}(0) = 0.16$.

The PL enhancement is fully reversible and can be deterministically controlled by switching on/off the MIR co-excitation. PL spectra were recorded over successive 20s on/off cycles of the MIR excitation, and the integrated intensity of the main emission peak at 436 nm was extracted. This data demonstrates a reproducible modulation of the emission intensity from the blue emitter ensemble during MIR co-irradiation, followed by recovery to the initial level when the MIR excitation is switched off (shown in Figure 2(c)). Notably, the PL response does not occur instantaneously; a finite delay is observed in both the onset and decay of the MIR-induced PL enhancement, indicating that longer time scales are at play in the MIR-modified dynamics of trapped carriers, as typically observed in blinking emitters.

To clarify whether the observed PL enhancement originates from changes in radiative or non-radiative decay processes, time-resolved lifetime measurements were carried out under MIR excitation and 405 nm pulsed laser at 30 MHz repetition rate and 50 μW. The lifetime comparison of the same blue emitter ensemble shows a slight change in the emitter lifetime by 0.04 ns from the initial value of 1.68 ns under MIR excitation (shown in Figure 2(d)). Other blue emitters exhibited only a slight reduction in lifetime of approximately 0.04 - 0.07 ns (shown in Figure SI1). These variations fall within the experimental error and therefore do not indicate a significant change in the emitter lifetime. This observation suggests that the MIR-induced PL enhancement does not arise from modifications of the intrinsic radiative or non-radiative recombination processes, but is more likely associated with changes in excitation efficiency or phonon-assisted population dynamics[5, 29, 30].

To investigate the wavelength selectivity of the MIR co-excitation and its influence on emitter response, the B centre was co-excited over a MIR wavelength range of 6.9 – 8.6 μm at a fixed MIR laser power of 60 mW. The MIR laser power was measured before the reflective objective lens, while the CW blue excitation laser power was maintained at 200 μW for all measurements. Figure 3(a) shows the PL spectra of ensembles of B centres across the MIR wavelength range with the spectrum recorded without MIR irradiation. Pronounced PL enhancements are observed under MIR co-excitation, particularly under MIR wavelengths ranging from 7.3 to 8.0 μm. This rather broad response is interpreted as a consequence of the localized nature of the phonon modes coupled to the SPE dynamics. For a more detailed analysis, the spectra were fitted using four Gaussian components, following the same method described above.

Figure 3(b) and figure SI2 summarise the fitted peak intensities and full width at half maximum (FWHM) values of the ZPL and the three phonon sidebands as a function of MIR wavelength compared to those without MIR irradiation. A pronounced enhancement in emission intensity is observed for the ZPL and the lower-energy phonon sidebands (PSB1 at 443 nm and PSB2 at 462 nm, respectively), whereas no significant change is observed for PSB3 at 480 nm. The MIR wavelength range of 7.3 – 8.0 μm closely overlaps with the in-plane infrared optically active mode of hBN (~1370 $cm^{-1}$). On the other hand, there's no significant PL enhancement at shorter MIR wavelengths despite their higher photon energy, as well as above 8.0 μm. This observation indicates that the enhancement mechanism arises from resonant phonon excitation, supporting the hypothesis that the MIR excitation selectively drives lattice vibrations coupled to the electronic state of the emitter. This is also consistent with recent reports on MIR-induced nonlinear optical enhancement in hBN[31].

Notably, the FWHM values of all emission features remain nearly constant across the investigated MIR wavelength range, indicating that the MIR-induced PL enhancement does not introduce measurable spectral broadening or additional dephasing mechanisms. Additionally, measurements performed on 30 emitters across hBN flakes with thickness ranging from 50 to 250 nm consistently showed reversible PL enhancement of 9 - 50% at the ZPL with a control of MIR laser excitation. Results from 20 representative emitters are shown in Figure 3(c). Each emitter was individually tested at a single MIR wavelength. Within the range of 6.9 – 7.8 μm, clear ZPL enhancement exceeding 10 % was consistently observed, with the largest enhancement achieving ~ 50% at 7.2 μm. Enhancements above 20 % were typically observed for MIR wavelength between 7.2 and 7.5 μm. This MIR wavelength selectivity is also consistent with the wavelength-dependent response measured from an ensembles in Figure 3(b).

Finally, figure 3(d) presents the MIR power-dependent PL response at a wavelength of 7.4 μm. The normalised ZPL intensity, defined as the ratio of the ZPL intensity under MIR co-excitation to that of the initial state (prior to MIR on state), is plotted as a function of MIR laser power. The PL enhancement reaches a maximum at an MIR power of approximately 25 mW and subsequently decreases with increasing power.

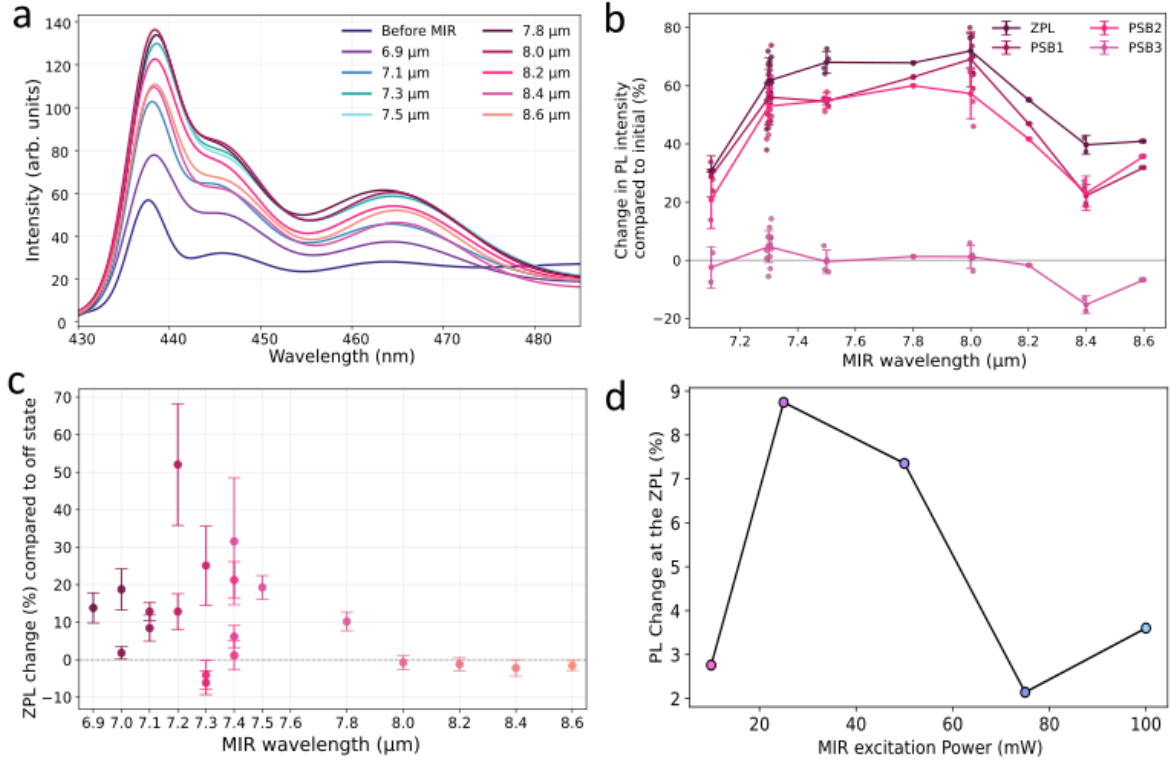

*Figure 3. Spectral analysis of a blue emitter ensembles under mid-IR excitation.* (a) PL spectrum fitted with four Gaussian peaks corresponding to the ZPL and PSBs for each MIR excitation wavelength. (b) ZPL and PSB intensities changes (%) as a function of mid-IR excitation wavelength. (c). **Relative change of ZPl intensity (%) as a function of MIR wavelength for 20 emitters. Each colored marker represents an individual emitter. Error bars show the measurement uncertainty for each emitter at a given wavelength.** (d) ZPL intensity change (%) as a function of MIR laser power compared to that with the MIR off state.

Figure 4(a) illustrates the proposed phonon-assisted optical enhancement mechanism. Under 405 nm laser excitation alone, radiative emission from the emitter is observed in the form of ZPL and broad PSBs. However, during the excitation cycle, electrons can be trapped in metastable states. When MIR co-excitation is introduced, the MIR photon energy is resonant with an MIR-active phonon mode of hBN ($\hbar\omega_{MIR} \approx \hbar\omega_{phonon}$), where $\omega_{MIR}$ is the MIR laser frequency and $\omega_{phonon}$ is the optical phonon mode frequency of hBN. A phonon-assisted excitation of the trapped carrier can overcome the trapping barrier, leading to an additional pathway to optically active states. This promotes the depopulation of trapped carriers and returns them to either the ground or the excited state. Thus, this process enhances the radiative emission without introducing measurable changes in the emission spectrum or linewidth. Importantly, time-resolved measurements reveal no significant change in the emitter lifetime, $\tau = 1/(\Gamma_{rad}+\Gamma_{non-rad})$, under MIR excitation, implying that the intrinsic radiative ($\Gamma_{rad}$) and non-radiative ($\Gamma_{non-rad}$) decay rates are preserved, which is consistent with this model when the metastable trap state lifetime is much longer than the radiative lifetime.

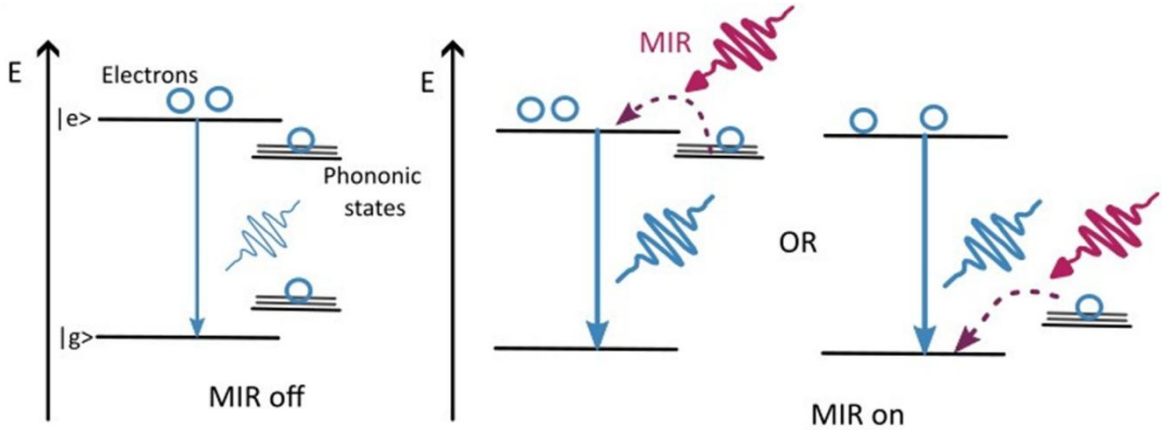

*Figure 4. Temperature-dependent photoluminescence measurements. Proposed illustration of MIR-induced phonon-assisted detrapping. The blue arrow indicates radiative relaxation pathways that give rise to the ZPL and associated PSBs. Filled blue circles and open blue circles represent holes and electrons, respectively. Under MIR co-excitation, the MIR field (red arrows) promotes phonon-assisted depopulation of carriers from metastable trap states, enabling the radiative cycle.*

In conclusion, we have demonstrated that a MIR laser can interact with solid state SPEs. Co-irradiation at the phonon frequency results in a reversible enhancement of the SPE emission. The intensity of the ZPL and associated PSBs were selectively increased when the MIR wavelength matched the in-plane optical phonon mode ($E_{1u}$) of hBN, while the emission FWHM, spectral profile, and lifetime remained unchanged. These experimental results establish MIR co-excitation as a reversible, non-destructive technique to optimise the optical performance of SPEs in solid-state materials through phonon-mediated control. It also opens an interesting pathway for interaction between phonon polaritons[32, 33] and quantum emitters in hBN.

**Methods**
**hBN exfoliation and emitter activation**
High-quality carbon-doped hBN crystals provided by the National Institute of Materials Science (NIMS) were mechanically exfoliated onto a CaF2 substrate using adhesive tape. The substrates were subsequently annealed on a hot plate at 400 °C for 3 hours to remove the tape residue, followed to in-situ air plasma cleaning inside the scanning electron microscope (SEM) chamber for 15 minutes to further eliminate the sample surface contamination. Emitter activation was carried by focused electron beam irradiation at an accelerating voltage of 3 keV and a beam current of 1.6 nA, with an electron dose of approximately $1\times10^{15}$ electrons per spot using patterning software to define 3*3 emitter arrays in the hBN flakes. After the electron beam irradiation, the samples were cleaned in a UV/Ozone cleaner for 2 hours to remove hydrocarbon-related contamination in the vicinity of the activated emitter. Flake thickness were measured using Park XE7 atomic force microscopy.

**PL characterisation**
All PL measurements were performed using a home-built confocal microscope system. A CW 405 nm laser was used for optical excitation and focused onto the sample through a 100x objective lens (Mitutoyo, M Plan Apo). A sample stage attached to the Nanocube piezo positionor was employed to construct PL maps. MIR source, a QCL with a power tuning of 6.9 to 8.6 µm (Daylight Solution, MIRcat) was spatially aligned with the 405nm excitation path. The MIR beam was focused onto the sample using a reflective microscope objective (Thorlab, LMM40X-P01), ensuring a consistent focal distance over the entire MIR tuning range. The beam spot size of the MIR laser on the sample surface is approximately 15 μm and the power density at 100 mW is about 0.260μW/μm$^2$. The MIR laser power was measured before the objective lens using a calibrated power meter. The detection path was split into two paths using a 50:50 beam splitter, enabling simultaneous photon counting with avalanche photodiodes (APDs) and spectral acquisition with a grating spectrometer (Ocean Optics, QE pro). PL maps were acquired using a 460 nm band-pass filter, while PL spectra were recorded using a 430 nm long-pass filter. The second-order autocorrelation measurements were performed using a time-correlated single photon counting module (PicoHarp300, PicoQuant), the signal was split evenly using a 50:50 beam splitter fibre in the collection path. The lifetime measurements were carried out using a 402 nm pulsed laser at 50 μW and 30 MHz repetition rate.

**Temperature-dependent PL**

Temperature-dependent PL measurements were conducted using a home-built heating stage powered by a DC supply. The sample temperature was monitored using a thermocouple attached to the end of the heating platform. After the sample temperature reached the set value and remained stable for at least 5 minutes, the PL spectra and maps were recorded.

**Acknowledgements**

The authors acknowledge Takashi Taniguchi (NIMS) for providing hBN crystals. The authors acknowledge financial support from the Australian Research Council (CE200100010, FT220100053), the Air Force Office of Scientific Research (FA2386-25-1-4044) and the Discovery Early Career Researcher Award (DE250100406). The research was funded by the Swiss National Science Foundation (SNSF project No. 214993).